# Label-free correlative morpho-chemical tomography of 3D kidney mesangial cells

Ankit Butola,[1,2,#,*] Biswajoy Ghosh,[1,#] Jaena Park,[2,3,4] Minsung Kwon,[2,3] Alejandro De la Cadena,[2,3] Sudipta S Mukherjee,[2,3] Rohit Bhargava,[2,3,4,5,6,7] Stephen A Boppart,[2,3,4,6,7,8 *], Krishna Agarwal[1, *]

[1]*Department of Physics and Technology, UiT The Arctic University of Norway, 9037 Tromsø, Norway*
[2]*NIH/NIBIB P41 Center for Label-free Imaging and Multiscale Biophotonics (CLIMB), University of Illinois Urbana-Champaign, 405 N. Mathews Ave., Urbana, IL 61801, USA*
[3]*Beckman Institute for Advanced Science and Technology, University of Illinois Urbana-Champaign, 405 N Mathews Avenue, Urbana, IL 61801, USA*
[4]*Department of Bioengineering, University of Illinois Urbana-Champaign, 1406 W Green Street, Urbana, IL 61801, USA*
[5]*Department of Chemical and Biomolecular Engineering, University of Illinois Urbana-Champaign, Urbana, Illinois*
[6]*Department of Electrical and Computer Engineering, University of Illinois Urbana-Champaign, Urbana, Illinois*
[7]*Cancer Center at Illinois, University of Illinois Urbana-Champaign, Urbana, Illinois.*
[8]*Carle Illinois College of Medicine, University of Illinois Urbana-Champaign, 506 S Mathews Avenue, Urbana, IL 61801, USA*
[#] *: Equal authorship*
*\*Email: ankitbutola321@gmail.com, krishna-agarwal@uit.no, boppart@illinois.edu*

Label-free characterization of biological specimens seeks to supplement existing imaging techniques and avoid the need for contrast agents that can disturb the native state of living samples. Conventional label-free optical imaging techniques are compatible with living samples but face challenges such as poor sectioning capability, fragmentary morphology, and lack chemical specific information. Here, we combined simultaneous label-free autofluorescence multi-harmonic (SLAM) microscopy and gradient light interference microscopy (GLIM) to extract both chemical specific and morphological tomography of 3D cultured kidney mesangial cells. Imaging 3D *in vitro* kidney models is essential to understand kidney function and pathology. Our correlative approach enables imaging and quantification of these cells to extract both morphology and chemical-specific signals that is crucial for understanding kidney function. In our approach, SLAM offers a nonlinear imaging platform with a single-excitation source to simultaneously acquire autofluorescence (FAD and NAD(P)H), second, and third harmonic signal from the 3D cultured cells. Complementarily, GLIM acquires high-contrast quantitative phase information to quantify structural changes in samples with thickness of up to 250 μm. Our correlative imaging results demonstrate a versatile and hassle-free platform for morpho-chemical cellular tomography to investigate functions such as metabolism and matrix deposition of kidney mesangial cells in 3D under controlled physiological conditions.

*Introduction:*

Label-free imaging seeks to enable direct observation of biological samples in their natural state, potentially enabling real time observations of biological processes with minimal perturbation. However, conventional bright field microscopy suffers from low intrinsic contrast that makes morphological imaging challenging and a lack of chemical specificity that makes observations of biological processes challenging. Emerging label-free imaging techniques seek to address some of these limitations– some in the realm of non-linear optics such as multiphoton autofluorescence and multiharmonic imaging[1] and the others in the realm of linear processes such as quantitative phase imaging (QPI), infrared[2] or Raman spectroscopic imaging[1,3,4], which all have their unique and complementary contrast mechanisms[5]. For example, QPI offers enhanced morphological imaging by utilizing an intrinsic optical contrast that encodes the thickness and local refractive index of the sample into phase of the light passing through the sample[4]. During the past few decades, variety of QPI techniques have been developed to improve their performance in terms of resolution, axial sectioning, and spatial and temporal phase sensitivity [6-8]. In general, it is challenging to choose the best existing QPI techniques due to multiple factors including resolution, acquisition speed, and particularly the suitability for specific applications[8-10]. For example, only a few existing QPI techniques can provide morphological information for samples with thicknesses exceeding 100 μm[11,12]. In addition, QPI lacks the chemical specificity required to distinguish features such as collagen, vesicles, and fibrosis as well as monitor physiologic processes that may not be manifest in morphologic changes[4,6,9]. This limitation hinders the potential impact of QPI in biomedical applications and prevents its use as a complementary tool for developing more selective diagnostic biomarkers.

On the other hand, label-free imaging with multi-photon excitation of autofluorescence and harmonic generation can complement QPI in terms of chemical specific imaging [13,14]. Multi-photon microscopy (MPM) relies on multi-photon excitation whereas, multi-harmonic microscopy (MHM) exploits the non-linear scattering of the incident light at the sample to image deep within tissue[15]. MPM encodes FAD and NAD(P)H coenzymes which provide insight into the chemical environments of cells and tissues such as oxidative and reductive chemical situations in cells and tissues[16]. In addition, MHM can visualize structural features such as actin, myosin, collagen fibers, extracellular vesicles, and water-lipid interfaces[13]. As individual modalities, each of these provides an interesting but narrow window into cells and tissues. For example, the quality of the multi-harmonic image depends on the spatial distribution of harmonophores, phase of the fundamental beams, and phase matching which may obfuscate quantitative interpretation of the resultant images[17]. In addition, multi-harmonic provides contrast from boundaries, and therefore encodes 2D morphology, but cannot indicate the optical thickness aspect of morphology.

Interestingly, the picture becomes more complete by performing correlative imaging using more than one imaging technique. Some strides have been made to support correlative multimodal label-free imaging, such as the simultaneous label-free autofluorescence-multi-harmonic (SLAM[18]) system and multimodal hyperspectral imaging systems[19]. However, integration with QPI systems has been quite challenging so far due to (a) use of the optical components from a different spectral window for both illumination and collection and differences in optical system design, (b) limited depth of penetration of QPI[4] and (c) difficulty in correlating volumetric morphological information imaged in QPI with single plane information encoded in non-linear label-free imaging modalities. Due to these challenges, the complementarity of QPI over the other techniques has not been significantly investigated. While gradient light interference microscopy (GLIM) solves problem (b), we here undertake the task to demonstrate (c) and therefore illustrate both the correlative and complementary aspects of QPI to multiphoton autofluorescence and multi-harmonic systems. The applicability to developed multimodality routes is shown on 3D cultured kidney mesangial cells. The kidney is a complex and vital organ, necessitating that changes need to be detected early to avoid long-term complications in diseases like glomerulonephritis and diabetic nephropathy. These diseases are closely associated with altered metabolism. With the advancement of 3D cell culture and related 3D biological models, it is now possible to simulate complex biological processes and measure response behaviors that are amenable to optical microscopy. These models mimic the intricate kidney microenvironment, allowing the study of cellular interactions, tissue organization, and disease mechanisms in a controlled setting. Measuring dynamic metabolic processes is important to identify early kidney disease markers and understand underlying events. Further, a morphological correlation that is quantitative in nature is important to understand spatial manifestations of the events.

In response to these challenges, we introduce a multimodal label-free imaging approach to reveal both morphology and chemical-specific information, and to measure the 3D nanoscale changes and cellular trafficking of cultured kidney mesangial primary cells. The nanoscale sensitivity of gradient light interference microscopy (GLIM)[12] is utilized to extract the morphological information of 3D mesangial cells. In addition, SLAM[18] microscopy supports simultaneous quantification of morphological, metabolic, and structural changes of samples with thickness of up to 250 μm over a large field of view. Therefore, correlative imaging using GLIM and SLAM allows nanometric morphological changes along the axial direction (with GLIM) and chemical specific imaging (with SLAM) at each plane of the sample. In this work, we demonstrate the capabilities of the system to extract morpho-chemical tomographic data of a 3D kidney organotype in a healthy state. 3D morphology of mesangial cell functions is investigated using GLIM and SLAM to quantify FAD, NAD(P)H, collagen, and optical heterogeneity from the mesangial cells to understand their functions such as metabolism, matrix deposition, and their roles in kidney pathology. The correlative approach offers a combination of functional and quantitative information of the cells with large depth of penetration, providing an imaging platform with a potential for high-throughput chemical and morphological imaging for specific biological applications.

*Experimental design and system configuration:*

Figure 1 shows a representative 3D cultured model of kidney mesangial cells that are responsible for crucial functions in the blood filtration process. High sugar concentrations during diabetic nephropathy affect the function of the mesangial cells[20]. The disease is manifested due to reasons like glucose levels, blood pressure, and inflammation which results in stiffening of the matrix and thereby leading to compromised function. Figure 1(a) illustrates the structural and functional components that can be exploited for understanding the functionality of mesangial cells. Figure 1(b) shows the 3D rendition of the model when the sample is fixed with a structural protein, actin, labelled. Figure 1(c) represents the schematic of multimodal label-free imaging approach to image both morphology and chemical-specific information to measure inter-cellular trafficking in cultured kidney mesangial cells. SLAM[18] and GLIM[12] system are shown in Fig. 1(c) which are utilized to extract nanoscale morphology and metabolic/structural changes of the sample, respectively. Note that, both modalities can provide morphological and metabolic changes of sample thickness up to 250 μm. Therefore, correlative imaging using GLIM and SLAM allows high-resolution morphological changes along axial direction (with GLIM) and chemical-specific imaging (with SLAM) at each plane of the sample. Further, the detectors are designed to collect a high signal-to-noise ratio signal from the sample to detect different functional aspects (Fig. 1d).

SLAM (20X, 1.05 NA) microscopy captured NAD(P)H, FAD autofluorescence, second harmonic, and third harmonic signals. A single excitation band across 1080-1140 nm is used to achieve simultaneous molecular contrast visualization in four detection channels. Near-transform-limited pulses with broader bandwidth enhance contrast, enabling a clearer interpretation of intercellular dynamics. SLAM combines 2-photon autofluorescence (2PAF), 3-photon autofluorescence (3PAF), second harmonic generation (SHG), and third harmonic generation (THG) imaging to extract FAD, NAD(P)H, collagen, and lipid-water interfaces from the kidney mesangial cells, respectively. In addition, GLIM (20X. 0.45 NA) is used to extract optical phase tomography of the cells. GLIM combines differential interference contrast (DIC) with a phase-shifting unit to extract the gradient phase at each layer of the sample. Selective interference of DIC and phase shifting suppress unwanted scattering for multiple layers of the samples and therefore help to produce tomography images of both thin (300 nm) and thick samples of up to 250 μm.

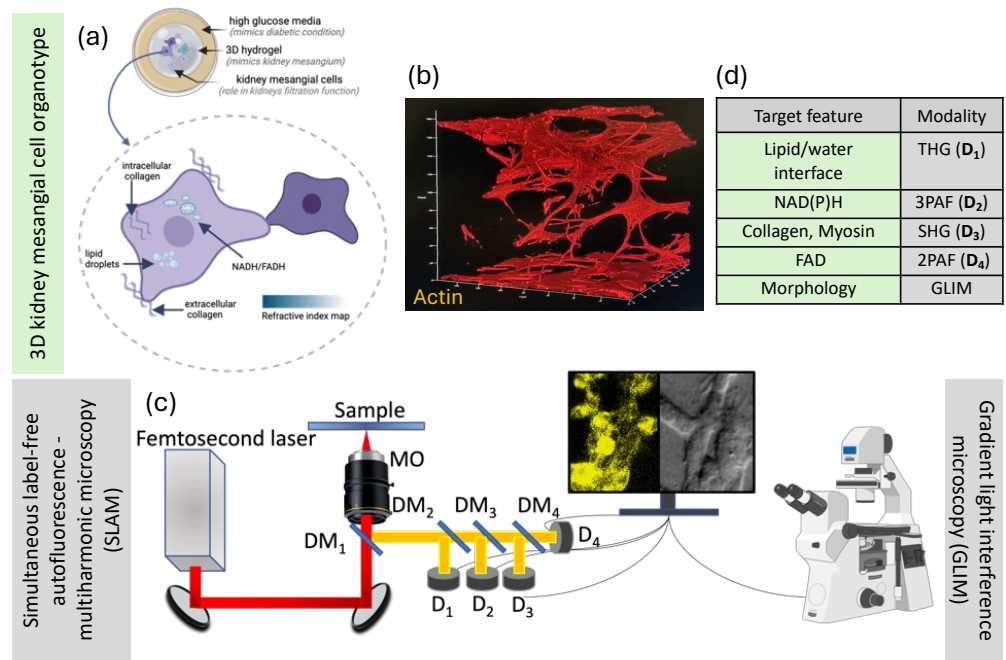

**Figure 1:** Illustration of the study design. (a) shows the configuration of the 3D organotypic of kidney mesangial microenvironment that dictates the renal filtration function. (b) 3D rendition of the model when the sample is fixed with actin labelled. The intracellular, extracellular, and cell-cell interactions determine the mesangial function and can be chemically imaged with (c) correlative SLAM (left) and GLIM (right) imaging setup. (d) Outlines the different aspects of the mesangial functions that can be targeted with the inverted imaging setup. D: Detector, DM: Dichroic mirror, MO: microscope objective.

*Sample preparation*:

A total of 12 samples with varying stiffness were prepared for the study. To prepare the samples, mesangial cells were commercially procured and grown in Dulbecco's Modified Eagle Medium (DMEM) with high glucose (4.5 g/l) supplemented with 20% fetal bovine serum (FBS) and 1% penicillin-streptomycin antibiotic. For the 3D cell culture, we used in-house prepared gelatin methacryloyl (GelMA). Total 2.5% (soft) and 5% (normal) (w/v) GelMA foam was dissolved in 1X phosphate buffered saline solution (PBS) and lithium phenyl-2,4,6-trimethylbenzoylphosphinate (LAP) was used as the photoinitiator. LAP was mixed with GelMA at a concentration of 5 mg/ml to make the precursor solution. GelMA was then cast on 35 mm glass bottom dishes and crosslinked with UV lamp exposure. Mesangial cells were in soft and normal stiffness GelMA hydrogels and allowed to grow for 96 hours.

*Results and discussion*:

Using an average power of 20 mW at the sample surface, SLAM images were acquired in a large field of view from a 3D kidney mesangial cell culture in soft and normal matrices *in vitro*. Total 12 different types of 3D cultured cells were visualized under the SLAM system. Multiple images were acquired from various region of interest (~25 FOV per sample). Each image was taken over 18 seconds. Different portions of these cells are identifiable based on their distinct features from the 4-channel-based optical signatures as shown in Fig. 2. Cells with localized 2PAF (cyan, FAD) and 3PAF (green, NAD(P)H) signals show cytoplasm, cells shapes, size, and nuclei as shown in Fig. 2 (a, b). In the mesangial cells that demonstrate a high metabolic state under diabetic conditions, the spatial distribution of 2PAF and 3PAF provides a window to measuring NAD(P)H and FAD. The ratio of Fig. 2 (a) and (b), which indicates the ratio of NAD(P)H and FAD, illustrates that the value can be greater or smaller than 1 depending upon the local density of the cells.

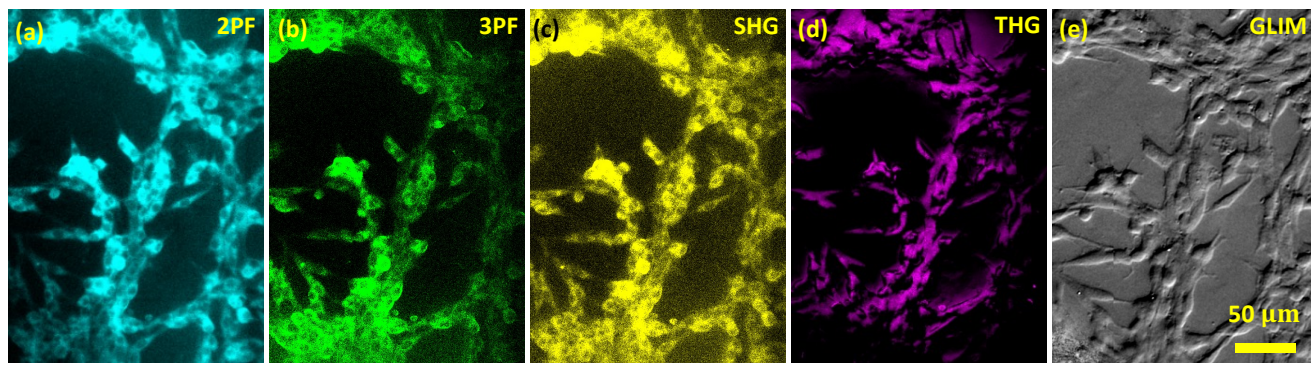

**Figure 2:** A comparative visualization of the mesangial organotype in the (a-e) SLAM and (g) GLIM imaging systems for the primary kidney mesangial cells. The GLIM images provide the gradient phase values of the cells which aid in identifying the content of the cells and their spatial distribution with the cells. The corresponding field of the SLAM images shows the metabolic and chemical profile of the cells including NAD(P)H, FAD, lipids, and fibrillar structures like actomyosin complex and collagen.

In addition to the shape and size of cells, our imaging of kidney mesangial cells in normal matrices *in vitro* (Fig. 2(c, d)) indicates overall structural changes, changes in the metabolic state, and cellular production of matrix proteins such as collagen, cytoskeletal structures, and lipid bodies. As collagen is produced by these cells in high glucose conditions, inflammation, and high stiffness, SHG (Fig. 2c) can be used to measure small changes in the development of fibrogenic manifestation via collagen accumulation. In addition, THG (Fig. 2d) signals primarily associated with the third order susceptibility and provide the contrast from vesicles and boundaries of the subcellular structures of the mesangium.

Further, SLAM findings are correlated with GLIM to understand the relationship between metabolism, fibrosis, and morphology, which are crucial for studying kidney disease *in vitro*. GLIM has a unique capability of suppressing the multiple scattering from the defocused layer of the thick samples. Partially coherence illumination in GLIM, selective interference of DIC, and four-phase shifting help to reject the multiple scattering, thus providing high-contrast imaging in the case of thick samples, which is unusual in other conventional phase imaging techniques. Figure 2(e) shows the GLIM image of kidney mesangial primary cells. Removing multiple scattering using GLIM helps to visualize intercellular details of complex 3D cellular systems. Figure 2(e) represents the derivative of the phase at each point of sliced portion of mesangial cells which shows the ability of GLIM to see through cellular layers that is well suited for other applications such as 3D tissue imaging. The ability to recover overall morphology and chemical-specific signals in thick samples hints at the potential of integrating these techniques in future use in *in vivo* applications.

To demonstrate the relevance of the correlative imaging, we investigated the cellular appendages or filopodia (Fig. 3). In cells these appendages can be as small as sub-micron to sub-diffraction sizes. Tunneling nanotubes (TNTs) are such filopodial extensions that are responsible for exchanging key organelles like mitochondria, vesicles, and nutrients between cells[21]. The exchange process is key for the physiological functioning of the cells and their alteration can be indicative of pathology. In kidney glomerular cells they are responsible for proper renal filtration function[22]. Such structures are replete with mitochondria, which means metabolic activity of NAD(P)H and FAD can be accessed. Further, the region will have actomyosin complexes for providing structure, as well as lipids from the cell membrane and lipid vesicles as transporters of various proteins.

Therefore, such a region has the potential for being investigated using multiphoton and multi-harmonic imaging. In Fig. 3a, we observe two different types of filopodia marked with white and yellow showing distinct phase gradients. This implies the dry mass of the white arrowed filopodia is higher. While 2-photon imaging showed the highest signal for the white arrowed filopodia, the white arrowed filopodia were sparingly visible (Fig 3b). Since 2-photon fluorescence captures FAD the filopodia may likely have a high density of mitochondria being transferred within. Figure 3f shows how a filopodium with high and almost no mitochondria looks like when labeled and imaged. In the 3-photon modality for imaging the NAD(P)H (Fig. 3c), although the filopodia are sparingly visible, the white arrowed one is brighter. Thus, using both 2PAF and 3PAF, the redox ratio of NAD(P)H/FAD can be obtained to measure the metabolic state. In SHG (Fig 3d), we again see very little signal from either of the filopodia, which is expected as the only source of non-centrosymmetric structure there is myosin and in such a thin structure the density is quite low. The samples were cultured for less than a week and hence substantial collagen was not deposited by the cells to be visualized by SHG. In the THG image (Fig 3e), we interestingly see both the filopodia are equally visible with nearly similar intensity. This is expected as in both structures there is only a single bilayer of lipid around the appendage, giving the structures equal weightage for signal.

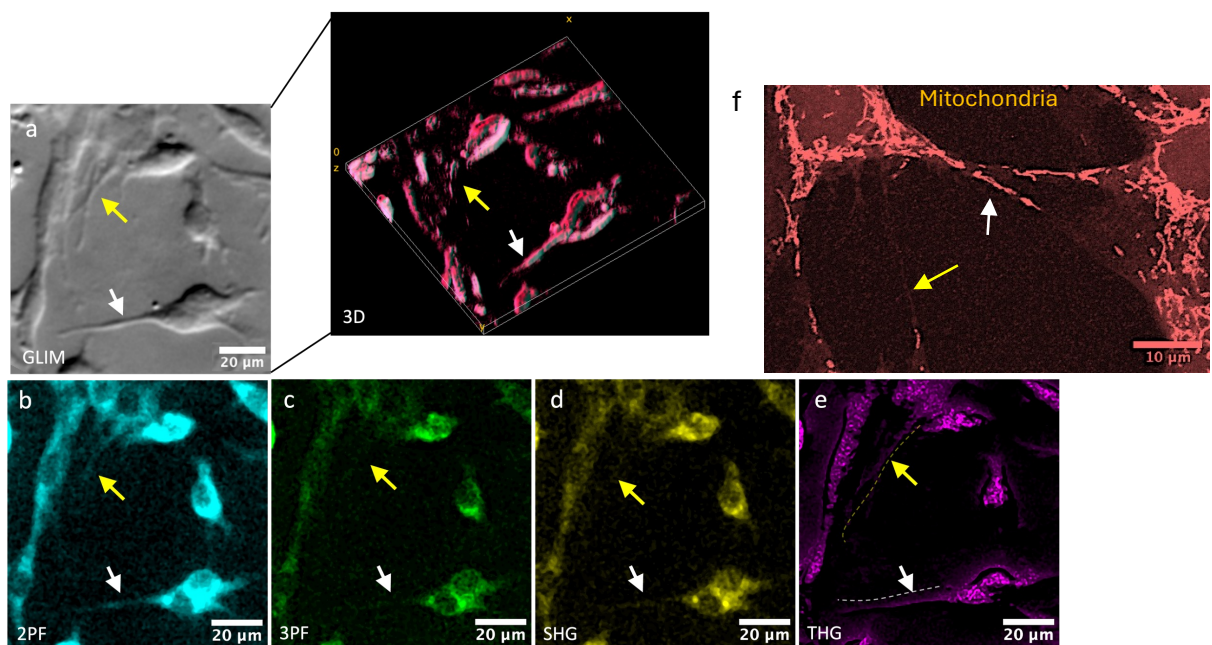

**Figure 3:** Figure illustrates cell-cell interaction among mesangial cells and the relevance of the SLAM-GLIM correlative imaging. (a) GLIM image and its 3D revisualization show the filopodia that function in coordinating activities in the mesangium. Two distinct filopodia are indicated by yellow and white arrows. (b) In 2PAF the white arrowed filopodia is very prominent, while the yellow arrowed is sparingly visible. (c) In 3PAF and (d) SHG the white arrowed filopodia is sparingly visible but the yellow arrowed is not visible at all. (e) The filopodia are visible in the THG image. (f) A fixed and labeled mitochondria region (same sample different ROI for representation), shows white-arrowed filopodia with several mitochondria being transferred, while yellow-arrowed filopodia is mostly empty and only one mitochondrion at its distal end.

Since different optical systems and magnification are used to acquire both SLAM and GLIM images, fiducial markers were used to locate the same ROI. However, subpixel matching of GLIM and SLAM images is challenging due to the different resolution, magnification, sample rotation, and mechanical vibrations of the different imaging systems. Nonetheless, the 2PAF, 3PAF, SHG, and THG images from the common ROI complement the GLIM image to extract metabolic and structural properties of the sample. Unlike marker-based microscopy techniques, the intensity at four different channels in SLAM microscopy is correlated with the concentration of auto-fluorophores. The separation between 2PAF and 3PAF signals indicates the potential of SLAM microscopy for *in-vivo* redox state imaging. Although correlative GLIM and SLAM imaging is performed to extract both linear and non-linear susceptibilities of the sample, our future aim will be to integrate these modalities to provide a common imaging platform for *in-vivo* live cell imaging.

### *Conclusion and future scope*

Here, we presented the importance of correlative morphological and chemical imaging to identify such clues from 3D biological organotypic models of the kidney mesangium. The kidney is a sensitive organ, and it is crucial to detect early pathological changes to avoid irreversible damage. Mesangium is a key microenvironment in the body that resides upstream of the disease development pathway. Thus, imaging such a structure will be crucial to identify early disease markers. We demonstrated that correlative label-free imaging methods can be useful to extract morphological and metabolic signatures from the 3D cultured cells. Although, more structural biological studies need to be planned, for example, through renal biopsies to address individual diseases like diabetic nephropathy where the diagnosis is very challenging as it silently affects the organ. Currently, the SLAM and GLIM setups are housed in different microscopes in this study. However, the study confirms the need for co-localized imaging which is crucial for identifying live cells and their sub-cellular organelles and dynamics at small scales ranging from sub-micron to the optical diffraction-limited resolution. As many of the chemical changes can be realized at high resolution, the need for balancing resolution and signal-to-noise ratio for 3D samples is a necessity. We also acknowledge the need for fast imaging as the cell-cell organelle transfers occur at significantly faster time scales[22].

**Funding.** Stadler Jacobsens forskningsfond, European research council (ERC) (804233), NIH/NIBIB Center for Label-free Imaging and Multiscale Biophotonics (CLIMB), 1P41EB03177. EU FET Open RIA project OrganVision (964800), and Digital Life Norway Cross Project Activity Grant 2023.

**Disclosure:** Authors declare no competing interests.